# Efficient non-collinear antiferromagnetic state switching induced by orbital Hall effect in chromium

*Hang Xie[1], Nan Zhang[1,2], Yuteng Ma[1,3], Xin Chen[1], Lin Ke[2], and Yihong Wu[1,3]\**

[1]Department of Electrical and Computer Engineering, National University of Singapore, Singapore 117583, Singapore

[2]Institute of Materials Research and Engineering, Agency for Science, Technology and Research, Singapore 138634, Singapore

[3]National University of Singapore (Chong Qing) Research Institute, Chongqing Liang Jiang New Area, Chongqing 401123, China

Recently orbital Hall current has attracted attention as an alternative method to switch the magnetization of ferromagnets. Here we present our findings on electrical switching of antiferromagnetic state in $Mn_3Sn$/Cr, where despite the much smaller spin Hall angle of Cr, the switching current density is comparable to heavy metal based heterostructures. On the other hand, the inverse process, *i.e.*, spin-to-charge conversion in Cr-based heterostructures is much less efficient than the Pt-based equivalents, as manifested in the almost one order of magnitude smaller terahertz emission intensity and spin current induced magnetoresistance in Cr-based structures. These results in combination with the slow decay of terahertz emission against Cr thickness (diffusion length of ~11 nm) suggest that the observed magnetic switching can be attributed to orbital current generation in Cr, followed by efficient conversion to spin current. Our work demonstrates the potential of light metals like Cr as an efficient orbital/spin current source for antiferromagnetic spintronics.

Spin Hall effect (SHE) has attracted significant interest in the last decade due to its ability to convert charge current to spin current, enabling efficient electrical manipulation of magnetization dynamics of ferromagnetic layer through spin-orbit torque (SOT).[1-4] The SHE arises from both intrinsic and extrinsic scattering processes in materials with sizable spin-orbit coupling (SOC), such as heavy metals (HMs). However, recent theoretical work unveils that, other than a transverse spin current from SHE, an electric field applied along the plane of a metallic film can also induce a transverse orbital current due to orbital Hall effect (OHE).[5-11] Unlike the SHE, the OHE does not require SOC and therefore it was predicted to exist in a wide range of materials including light metals such as Al, Cu, and Cr.[9-11] Although the OHE is perceived to be more fundamental than SHE, experimental investigations on OHE are relatively recent because of primarily two reasons – difficulty in separating the two effects that typically occur concurrently in materials with the SOC and absence of direct coupling between orbital Hall current and magnetization. Nevertheless, there is increasing evidence that the OHE is present in various light metals with negligible or weak SOC,[12-15] including the orbital Rashba-Edelstein effect.[16-18] Similar to SHE, the OHE can also induce SOT,[12-14,16] orbital Hall magnetoresistance,[17] and unidirectional orbital Hall magnetoresistance[19] in ferromagnet (FM)/non-magnetic metal (NM) bilayers. Notably, SOT-induced magnetization switching has been demonstrated in CoFeB/Cr without a heavy metal layer.[13] In view of successful observation of OHE-related phenomena in HM/NM bilayers, naturally it would be of interest to understand how the orbital Hall current interacts with antiferromagnets (AFMs), which so far remains unexplored.

Among the AFMs, non-collinear AFMs, such as $Mn_3Sn$ and $Mn_3Ge$, have attracted great attention recently due to their large anomalous Hall,[20-23] anomalous Nernst,[24] and magneto optical Kerr effects,[25] which help to overcome the electrical readout difficulty faced by conventional collinear AFMs. A crucial step towards device application of non-collinear

AFMs is the realization of electrical control of the AFM magnetic order, which has been demonstrated experimentally by several groups, ranging from partial to full switching and continuous rotation of the chiral spin structure.[26-32] While most studies employ AFM/HM bilayer structures in which spin currents are generated from adjacent HM layers,[26-30,32] recent studies have shown that the AFM order of polycrystalline $Mn_3Sn$ can be switched by self-generated spin-polarized currents.[31,33-35] Despite these progresses, the switching mechanism remains partially understood, such as the role of crystalline orientation[27,30,31] and Joule heating.[28,29] To further understand the switching mechanism and thus reduce the switching current, it is useful to explore other sources of spin current besides SHE-generated or self-induced spin current.

Here we present a study of electrical control of AFM state in $Mn_3Sn$/Cr bilayers, where Cr is a light metal reported to have a small SOC but a large orbital Hall conductivity (OHC).[11,36,37] We observe efficient current-induced switching in $Mn_3Sn$/Cr with the switching polarity same as $Mn_3Sn$/Ta but opposite to $Mn_3Sn$/Pt, and the switching current density $J_c$ is comparable to $Mn_3Sn$/Ta but less than half of $Mn_3Sn$/Pt. The thickness and temperature dependence study of current-induced switching further rules out the effects from interfaces and antiferromagnetic order in Cr, respectively. Similar current-induced switching results are obtained in Cr/CoFeB and Ta/CoFeB control samples, both of which exhibit same switching polarity. We further perform terahertz (THz) emission measurements in CoFeB/Cr and CoFeB/Pt and find that the THz intensity in CoFeB/Cr is much weaker than that of CoFeB/Pt under same pumping conditions. From the Cr-thickness dependence, we obtain a spin (or orbital) diffusion length of 11 nm in Cr. These results combined suggest that the current induced switching in $Mn_3Sn$/Cr and Cr/CoFeB is dominated by the OHE rather than SHE based mechanism.

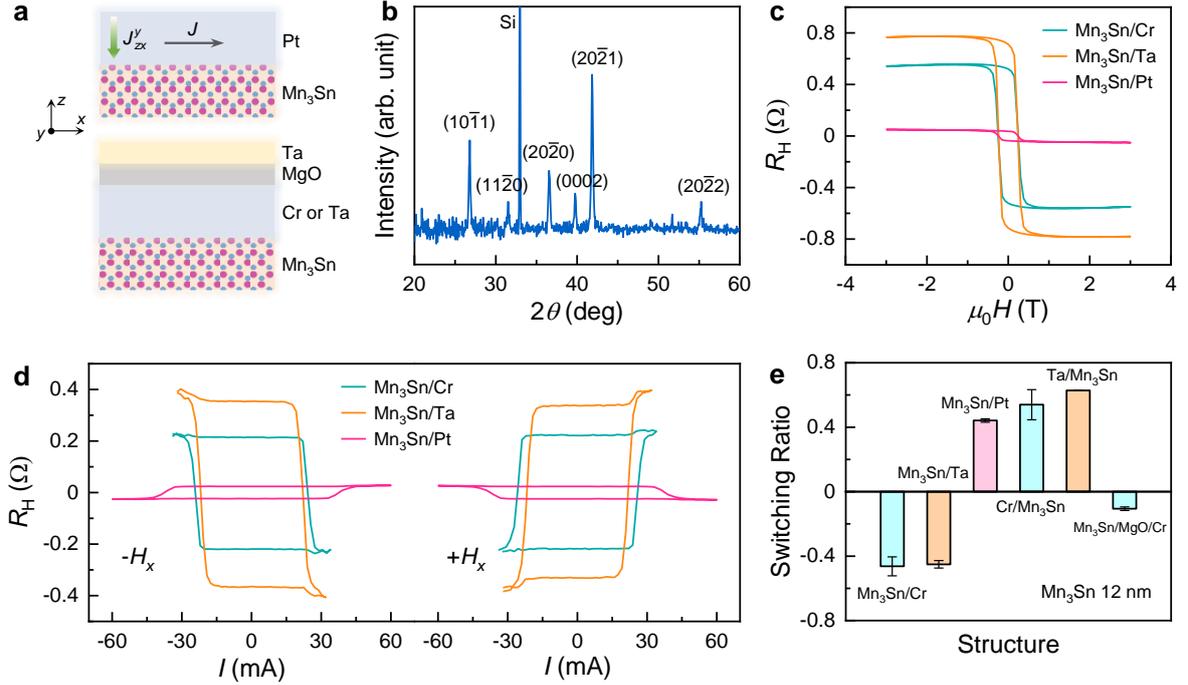

**Figure 1.** Structural properties of Mn$_3$Sn and current-induced switching in Mn$_3$Sn/X, X = Cr, Ta, Pt. **a,** Schematic layer structure of sample investigated. For Mn$_3$Sn/Cr and Mn$_3$Sn/Ta structures, additional MgO(2)/Ta(1.5) capping layers (number inside the parentheses indicate thickness in nm) are used to protect the film from oxidation. The thin films are patterned into Hall bars for electrical measurements. **b,** XRD pattern of a Mn$_3$Sn(60) coupon film. **c,** Field dependence of Hall resistance at room temperature for Mn$_3$Sn(12)/X(6) with X = Cr, Ta, Pt. **d,** Current dependence of Hall resistance at room temperature for Mn$_3$Sn(12)/X(6) with X = Cr, Ta, Pt, with a negative (left) and positive (right) in-plane assistant field. The current pulse width is fixed as 5 ms. **e,** Summary of switching ratio (the ratio between current and field induced Hall resistance change) of Mn$_3$Sn in different structures. The sign represents the switching polarity.

The schematic layer structure is shown in Figure 1a. The Mn$_3$Sn thin films are prepared using magnetron sputtering (see Supporting Information S1 for detailed methods). The non-magnetic metal (NM) layer X (X = Cr, Ta, Pt) is deposited after the annealing of Mn$_3$Sn to avoid possible interdiffusion (see Supporting Information S2). Figure 1b shows the X-ray diffraction (XRD) pattern of a $Mn_3Sn(60)$ film. Multiple peaks corresponding to $(10\bar{1}1)$, $(11\bar{2}0)$, $(20\bar{2}0)$, $(0002)$, $(20\bar{2}1)$, and $(20\bar{2}2)$ planes of $D0_{19}$ phase $Mn_3Sn$ are observed, confirming the polycrystalline structure in sputter-deposited $Mn_3Sn$. As shown in Figure 1c, all $Mn_3Sn(12)/X(6)$ films exhibit clear AHE hysteresis loops with similar shape and coercivity, although the absolute magnitudes of AHE resistance differ from sample to sample due to different current shunting effects. The results suggest that the X layer on top of $Mn_3Sn$ does not have much effect on the magnetic properties of $Mn_3Sn$.

Current-induced switching experiments are then performed using current pulses and an in-plane assistant field $H_x$ along the current direction (see Supporting Information S5 for more details on $H_x$ dependence). Deterministic switching behaviour is clearly observed in all the structures (Figure 1d). The switching polarity of $Mn_3Sn/Ta$ and $Mn_3Sn/Pt$ is consistent with the reported results and corroborates the negative and positive spin Hall angle (SHA) of Ta and Pt, respectively.[26,30,31] Despite opposite switching polarity, both show similar switching ratio of around 45% as summarized in Figure 1e. Surprisingly, $Mn_3Sn/Cr$ also shows evident switching loops with comparable switching ratio and critical current as $Mn_3Sn$/HM bilayers. We can also notice that its switching polarity is the same as that of $Mn_3Sn/Ta$, implying a negative SHA in Cr. We also reverse the stack deposition sequence of $Mn_3Sn/Cr$ (see Supporting Information S3). The extracted switching ratio of $Cr/Mn_3Sn$ together with the value of $Ta/Mn_3Sn$ is also summarized in Figure 1e. As can be seen, similar to $Ta/Mn_3Sn$, $Cr/Mn_3Sn$ has an opposite switching polarity with $Mn_3Sn/Cr$, whose switching ratio also increases. As standalone $Mn_3Sn$ also shows self-induced switching behaviour (with switching ratio usually smaller than 20%), and moreover, the self-induced switching polarity of $Mn_3Sn$ can vary depending on the underlayer due to the change in crystalline structure,[31] we further added a thin MgO layer between $Mn_3Sn$ and Cr layers in $Mn_3Sn/Cr$ to exclude dominant contribution from self-induced switching (Figure 1e and Supporting Information S4). The much-reduced switching ratio after the MgO layer insertion indicates that the switching in $Mn_3Sn/Cr$ is mainly induced by the external orbital current from Cr layer.

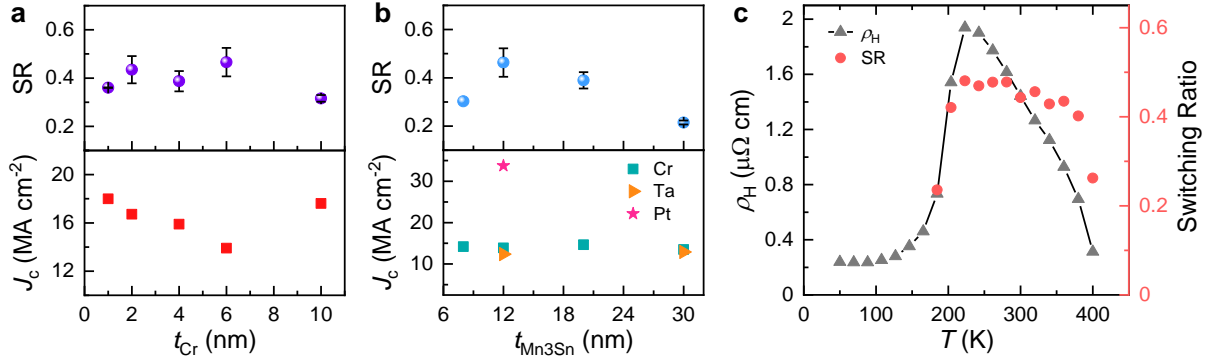

**Figure 2.** Cr thickness, $Mn_3Sn$ thickness, and temperature dependence of current-induced switching in $Mn_3Sn/Cr$. **a,** Cr thickness $t_{Cr}$ dependence of switching ratio and critical current density $J_c$ in $Mn_3Sn/Cr$, with $Mn_3Sn$ thickness fixed as 12 nm. The current shunting effect has been taken into consideration based on the resistivity of each layer for the estimation of current density. **b,** $Mn_3Sn$ thickness $t_{Mn3Sn}$ dependence of switching ratio and $J_c$ in $Mn_3Sn/Cr$, with Cr thickness fixed as 6 nm. **c,** Anomalous Hall resistivity $\rho_H$ and switching ratio of $Mn_3Sn(12)/Cr(6)$ as a function of temperature.

To understand the mechanism of current-induced switching in $Mn_3Sn/Cr$, we further study the thickness dependence of current-induced switching in $Mn_3Sn/Cr$ by varying the thicknesses of both Cr and $Mn_3Sn$ layers. The detailed AHE and current-induced switching loops are shown in Supporting Information S6 and S7, from which the extracted switching ratio and critical current density $J_c$ in Cr layer are plotted in Figure 2a and 2b, respectively. With varying Cr thickness $t_{Cr}$ and fixing $Mn_3Sn$ thickness $t_{Mn3Sn}$ at 12 nm, the switching ratio overall is around 40%, while $J_c$ firstly decreases from 18 MA cm$^{-2}$ to 13.9 MA cm$^{-2}$ when $t_{Cr}$ increases from 1 nm to 6 nm and then increases back to 17.6 MA cm$^{-2}$ at $t_{Cr}$ = 10 nm (Figure 2a). The initial decrease of switching current density can be explained by the increased amount of spin current generation as $t_{Cr}$ increases, whereas the increase at 10 nm might be due to onset of self-current shunting effect when $t_{Cr}$ exceeds a certain critical thickness (equivalent to spin diffusion length in HM). The switching ratio has a larger variation when $t_{Mn3Sn}$ varies– increases from 30% to 46% from 8 nm to 12 nm and afterwards keeps decreasing to around 21.5% at $t_{Mn3Sn}$ = 30 nm (Figure 2b, top). The switching ratio for polycrystalline Mn$_3$Sn is highly sensitive to the amount and spatial distribution of crystallites with different orientations, as the switching mainly occurs when the kagome plane is parallel to the spin polarization direction.[27,30,31] Therefore, it can vary largely with the Mn$_3$Sn thickness and even among samples with the same structure, due to the difference in detailed crystalline orientations. The $J_c$, however, shows very small variation with $t_{Mn3Sn}$ (Figure 2b, bottom), which is distinct from the results for FM/HM bilayer where $J_c$ increases as the magnetic layer thickness increases due to more magnetic moments needed to be switched. This may be partly caused by the heating effect as when the Mn$_3$Sn thickness increases more current flows in the Mn$_3$Sn layer, resulting in an increase of temperature in Mn$_3$Sn and suppression of $J_c$ increase at large Mn$_3$Sn thickness.[28,29] But, as discussed in Supporting Information S8, Joule heating should play a minor role in the present case. Nevertheless, by comparing the $J_c$ of Mn$_3$Sn/Cr with the values for Mn$_3$Sn/Ta and Mn$_3$Sn/Pt, we can find that $J_c$ of Mn$_3$Sn/Cr is merely slightly larger than that of Mn$_3$Sn/Ta and less than half of the Mn$_3$Sn/Pt bilayer, indicating that Cr can induce as efficient switching in Mn$_3$Sn as the HM does. However, this cannot be interpreted by the conventional SHE scenario given that Cr is a 3d light metal. Furthermore, interface-related effects such as Rashba-Edelstein effect (REE) also fail to account for it, because larger switching efficiency is not observed at either smaller Cr or Mn$_3$Sn thickness that would lead to less current shunting effect and thus favours interfacial effect.

In addition, Cr itself exhibits an AFM ordering with a Néel temperature $T_N$ of 311 K, which may give rise to specific spin transport phenomenon.[36,38] Although $T_N$ for thin film Cr can be well below room temperature, it is still prudent to vary the measurement temperature across $T_N$ to investigate if the AFM order in Cr is relevant to the spin current generation. It is found that the switching ratio displays a plateau-like shape as a function of temperature and maintains a level above 40% from 204 K to 380 K (Figure 2c and Supporting Information S9). Clear deterministic switching with a switching ratio around 26% is observed at 400 K that is much higher than $T_N$ of bulk Cr. The sharp drop below 204 K and above 380 K can be attributed to the disappearance of chiral AFM spin structure below the transition temperature and above Néel temperature of $Mn_3Sn$[39,40] (Figure 2c, $\rho_H$). Therefore, the AFM ordering in Cr is unlikely to play a role in the spin current generation in $Mn_3Sn$/Cr bilayers.

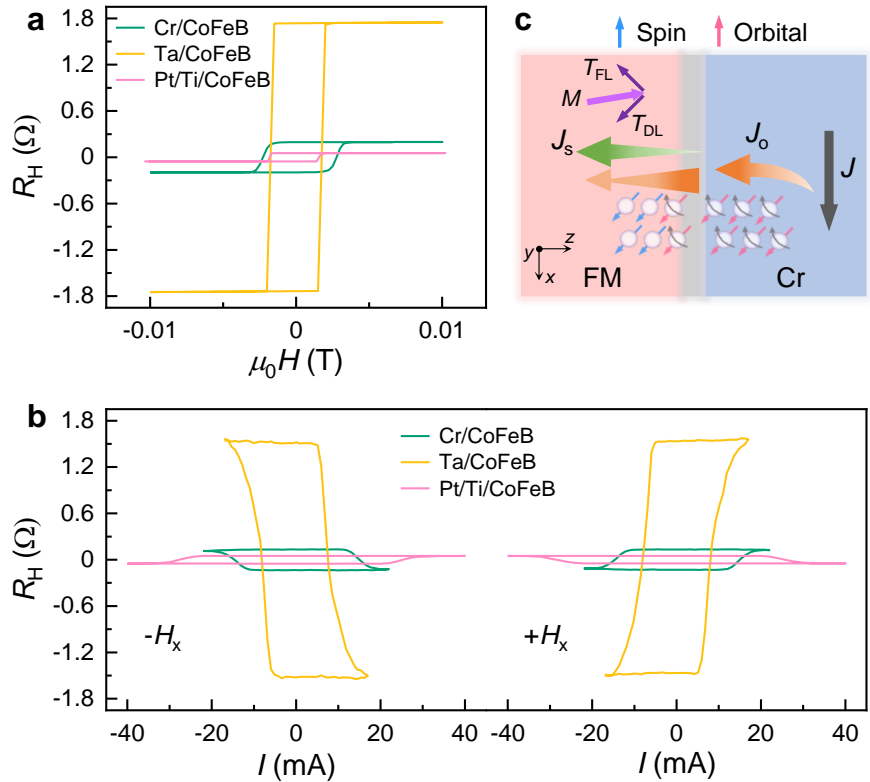

**Figure 3.** AHE and current-induced switching in Cr/CoFeB, Ta/CoFeB, and Pt/Ti/CoFeB. **a,** Field dependence of Hall resistance of Cr(6)/CoFeB(1.2), Ta(6)/CoFeB(1.5), and Pt(6)/Ti(2)/CoFeB(1.2). **b,** Current dependence of Hall resistance of Cr(6)/CoFeB(1.2),

Ta(6)/CoFeB(1.5), and Pt(6)/Ti(2)/CoFeB(1.2) with negative (left) and positive (right) in-plane assistant field. **c,** Schematic illustration for the OHE-induced spin current in the FM/Cr bilayers. Here, FM includes both $Mn_3Sn$ and CoFeB, $J_s$ and $J_o$ are spin current and orbital current.

To examine if such phenomenon exists only in non-collinear AFMs, we further test FM-based control samples, Cr/CoFeB, Ta/CoFeB, and Pt/Ti/CoFeB, where ultrathin CoFeB layer with perpendicular magnetic anisotropy (PMA) is employed for current-induced switching measurements. The thickness of CoFeB is tailored for formation of good PMA (Figure 3a and Supporting Information S10). Since CoFeB directly grown on Pt does not show PMA, a 2 nm Ti layer is added under CoFeB for PMA formation. As shown in Figure 3b, all three structures show clear deterministic switching loops driven by applied charge current. Cr(6)/CoFeB(1.2) has the same switching polarity as Ta(6)/CoFeB(1.5) and opposite switching polarity to Pt(6)/Ti(2)/CoFeB(1.2), which is consistent with the $Mn_3Sn$ case, indicating the spin torque switching in $Mn_3Sn$/Cr is rooted from Cr and can also apply to normal FM layer.

Observation of large SHA via inverse SHE voltage measurements in some 3d transition metals with small atomic number has been reported before, which was attributed to orbital filling of *d*-electrons.[38,41] It is predicted that in addition to the spin and charge degrees of freedom, the transition metals also have orbital degrees of freedom and can generate an orbital current in a nonequilibrium state due to OHE.[5-8] Due to the weak dependence of OHC on spin-orbit coupling, even some 3d, 4d metals and their related oxides exhibit giant values of OHC.[5,6,8,11] Therefore, we believe the current-induced switching in $Mn_3Sn$/Cr can be described in a scenario of OHE from Cr. As depicted in Figure 3c, when a charge current is applied along the in-plane direction, a transverse orbital current is generated due to the OHE.

This orbital current then propagates towards the magnetic layer and is further converted to a secondary spin current at both the interface and the interior of FM via the spin-orbit coupling $\langle \mathbf{L} \cdot \mathbf{S} \rangle_{FM}$ of the magnet. Therefore, different with the SHE situation, the strength and polarization direction of the resultant spin current also rely on the sign of $\langle \mathbf{L} \cdot \mathbf{S} \rangle_{FM}$. It was previously believed that the SOC in 3d ferromagnets is weak, while recently there are increasing evidences of sizable SOT and SHA from 3d FM and AFM thin films such as Co, Fe, Ni, CoFeB and FeMn.[42-47] Additionally, as aforementioned, spin current generation in $Mn_3Sn$ has also been reported both theoretically and experimentally, which arises from non-collinear spin structure with SOC.[22,31,33-35,48-52] These prerequisites allow orbital Hall current to efficiently converts to spin current when it enters $Mn_3Sn$ or CoFeB. Regarding the polarization direction of the resultant spin current from Cr, our results are consistent with the reported results in YIG/Cr,[38,41] Co/Cr,[13,14] CoFeB/Cr,[13] but opposite with the direction of Ni/Cr,[13,14] which we consider to be related to the SOC sign in the magnetic layer because it is given by the product of the OHC sign in Cr and the SOC sign in magnetic layer, for example, the SHA is reported to be negative in Co and Fe, but positive in Ni.[42] For $Mn_3Sn$, its sign of SOC is still unclear. Based on our result and the positive OHC in Cr, it can be inferred that it is negative.

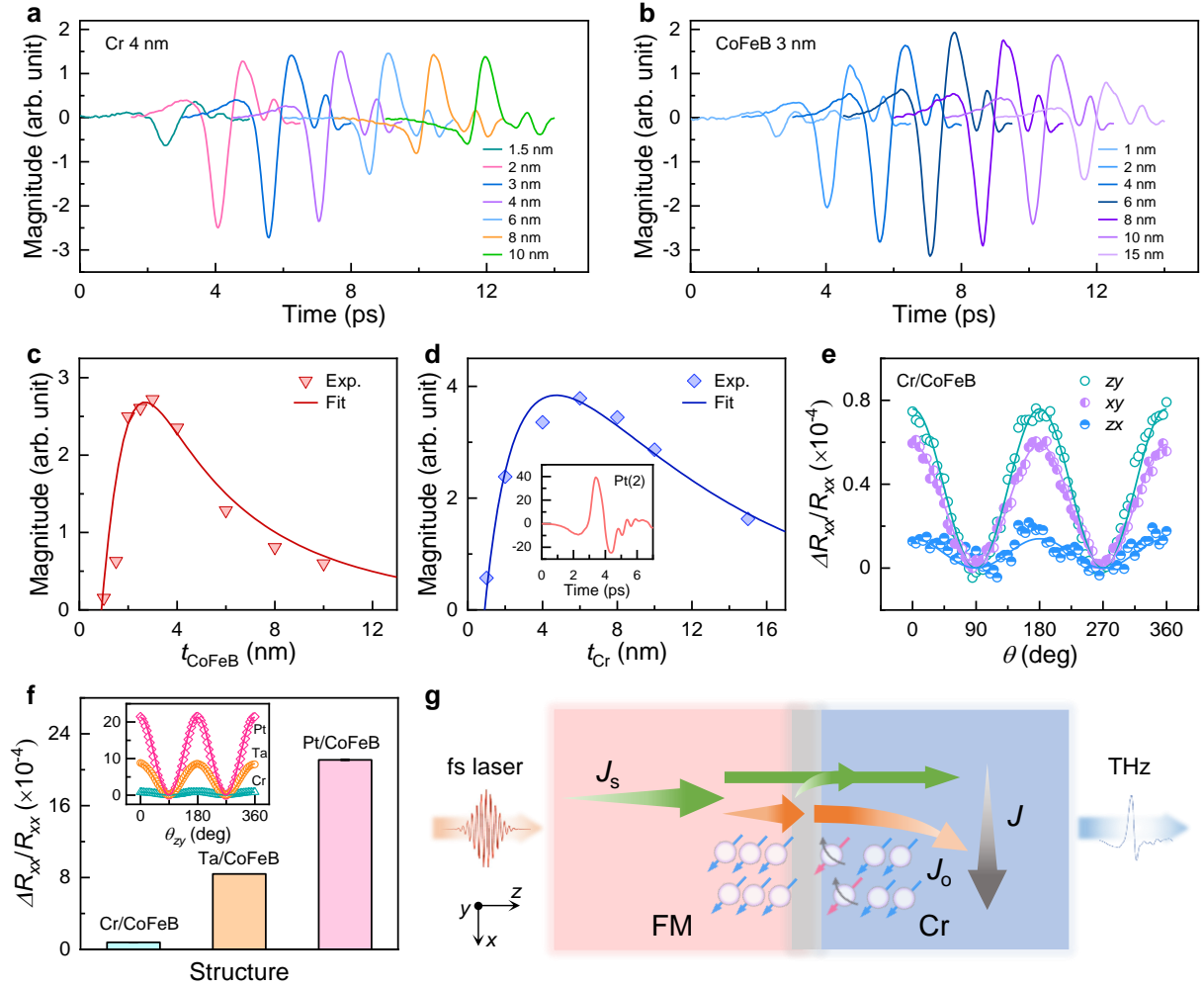

**Figure 4.** Cr thickness and CoFeB thickness dependence of THz waveform from CoFeB/Cr. **a**, THz waveform from CoFeB/Cr at different CoFeB thickness $t_{CFB}$, with fixed Cr thickness at 4 nm. Waveforms are shifted in time axis for clarity. A magnetic field is applied to the sample in measurement to align the magnetization along +*y* direction. **b**, THz waveform from CoFeB/Cr at different Cr thickness $t_{Cr}$, with fixed CoFeB thickness at 3 nm. **c,** Extracted THz magnitude as a function of $t_{CFB}$ (triangular symbol) and the fitting (solid line). **d,** Extracted THz magnitude as a function of $t_{Cr}$ (diamond symbol) and the fitting (solid line). The inset in **d** shows THz waveform from CoFeB(3)/Pt(2). **e**, Angular dependent magnetoresistance (symbols) in Cr(6)/CoFeB(1.2) at current amplitude of 2 mA and magnetic field of 3 T with field rotated in *zy, xy, zx* planes. $\theta = 0$ is along *z*-axis for *zy* and *zx* scan and along *x*-axis for *xy* scan. Solid lines are fittings. **f**, Summary of extracted MR($\theta_{zy}$) in Cr(6)/CoFeB(1.2),

Ta(6)/CoFeB(1.5), and Pt(6)/CoFeB(1.2). The inset in **f** shows detailed angular dependent MR curves of (Cr, Ta, Pt)/CoFeB. **g**, Schematic of THz generation in FM/Cr bilayers, which includes spin-to-orbital current conversion process.

To gain further insights of the charge-to-spin conversion mechanism in Cr-based structures, we examine its inverse process, *i.e.,* spin-to-charge conversion, through THz emission measurements. As the THz emission from Mn$_3$Sn is too weak to detect, possibly due to non-uniform magnetization and spin current polarization directions, we only focus on CoFeB/Cr bilayers. Figure 4a and 4b show the THz emission waveforms of CoFeB($t_{CFB}$)/Cr($t_{Cr}$) with varied $t_{CFB}$ and $t_{Cr}$, respectively. Clear THz emission has been observed throughout the thickness range with the polarity opposite to that of CoFeB/Pt, which is consistent with the electrical measurement results. Figure 4c and 4d show the corresponding $t_{CFB}$ and $t_{Cr}$ dependence of THz intensity. As can be seen in Figure 4c, the THz intensity initially increases from 1 nm to 3 nm, after which it decreases monotonically as $t_{CFB}$ further increases. The thickness dependence can be fitted using[53,54]

$$E_{THz} \propto \frac{A}{t_{CFB}+t_{Cr}} \cdot \frac{\tanh\left(\frac{t_{CFB}-t_0}{2\lambda_{CFB}}\right) \cdot \tanh\left(\frac{t_{Cr}-t_1}{2\lambda_{Cr}}\right)}{n_{air}+n_{sub}+Z_0(\sigma_{CFB}t_{CFB}+\sigma_{Cr}t_{Cr})} \cdot e^{-\frac{(t_{CFB}+t_{Cr})}{s_{THz}}}, \qquad (1)$$

where $A$ is the absorptance of the metal layers, $t_0$ is critical thickness of CoFeB layer, below which CoFeB layer may not show ferromagnetic properties, $t_1$ is the critical thickness of Cr layer, $\lambda_{CFB}$ is the spin dephasing length in CoFeB, $\lambda_{Cr}$ is the orbital diffusion length in Cr, $n_{air}$ ($n_{sub}$) is the refractive index of air (substrate), $Z_0$ is the impedance of free space, $\sigma_{CFB}$ and $\sigma_{Cr}$ are the conductivities, and $s_{THz}$ is the effective inverse attenuation coefficient of THz radiation in the bilayers. The values of $n_{air}$, $n_{sub}$, $Z_0$, $\sigma_{CFB}$, $\sigma_{Cr}$ are known: $n_{air} = 1$, $n_{sub} = 1.453$, $Z_0 = 377$ Ω, $\sigma_{CFB} = 6.99\times10^5$ S·m$^{-1}$, $\sigma_{Cr} = 7.87\times10^5$ S·m$^{-1}$, $t_{Cr} = 4$ nm. By further using the following parameters $t_0 = 0.9$ nm, $\lambda_{CFB} = 0.8$ nm, and $s_{THz} = 16$ nm, we could fit the thickness dependence well using Eq. (1) (Figure 4c, solid line). The fitted values of $t_0$, $\lambda_{CFB}$, and $s_{THz}$ are in good agreement with the reported value in FM/NM bilayers.[54] Similarly, by fitting the amplitude as a function of $t_{Cr}$ using Eq. (1) (Figure 4d, solid line), we obtain the value of $\lambda_{Cr} =$ 11 nm. This is comparable with reported value for Cr[41] and much larger than the spin diffusion length in HM layer, which is usually around 1-2 nm.[53,54] Besides, by comparison with the result of CoFeB/Pt bilayers (Figure 4d, inset), we find that THz amplitude in CoFeB/Cr is only 10% of CoFeB/Pt, implying much less effective spin-to-charge conversion process in CoFeB/Cr. The much weaker THz signal in CoFeB/Cr than CoFeB/HM was also reported in a previous study.[53]

The above THz results show that despite the efficient charge-to-spin conversion in current switching experiments, FM/Cr fares much worse in the inverse process, *i.e.*, spin-to-charge conversion, compared to FM/(Pt, Ta). This is further supported by the angular dependent magnetoresistance (MR) measurements. As displayed in Figure 4e, the MR of Cr(6)/CoFeB(1.2) shows $\cos^2\theta$ relationship with $\theta$ for all three measurement configurations (see figure caption for $\theta$ definition). The angle-dependence in *zy* and *xy* scan agrees well with those of conventional FM/HM bilayers, while the MR in *zx*-scan shows an opposite sign (with its origin unclear at present). The MR obtained in the *zy*-scan corresponds to the spin Hall magnetoresistance (SMR) in FM/HM bilayers, but considering its different origin, here we call it orbital Hall MR (OMR). The extracted OMR ratio in Cr/CoFeB is around $0.79 \times 10^{-4}$, which is more than one order of magnitude smaller than the SMR ratio in Ta/CoFeB and Pt/CoFeB (Figure 4f and inset). The THz and OMR results combined suggest that SHE is not the dominant mechanism responsible for the spin current generation in FM/Cr, because otherwise, one would expect a strong THz emission and large SMR from CoFeB/Cr, as with conventional FM/HM bilayers. The sharp contrast between current-induced switching and THz/OMR results may be understood based on the OHE scenario. As aforementioned, in current-induced switching, the OHE-induced orbital current needs to be converted to spin current at the FM/Cr interface and/or inside the FM to act on the FM layer (Figure 3c). The conversion takes place within a finite region in the FM layer near the interface, which may not affect the switching process as the torque can be generated "onsite" along with the orbital-to-spin conversion. However, this spatial separation from the generation sites of orbital current would significantly weaken the inverse process as depicted in Figure 4g. In the case of THz experiments, the inverse process, *i.e.*, spin-to-orbital conversion, is understandably inefficient due the fast speed ($v$) of laser-pumping generated hot electrons (as a rule of thumb, in the case of extrinsic process, the scattering cross section is proportional to $v^{-4}$). This simple fact explains why the THz emission is weak in CoFeB/Cr bilayers. On the other hand, unlike the spin current that is directly reflected at the interface in FM/HM bilayer, the orbital current needs to experience extra steps of orbital-spin conversion before it can be reflected from or modulated by the magnetization direction of the FM layer and, furthermore, the whole process takes place within a finite thickness region instead of right at the interface, and therefore, the eventual reflected orbital current will be much smaller compared to spin current reflection, leading to diminishing OMR. It is worth noting that large inverse conversion efficiency has been observed previously in Cr-based ferromagnetic heterostructures in spin-pumping experiments,[38,41] which differs from both the THz and OMR measurements in this work, as near thermal equilibrium spins are directly injected into the Cr layer and a significant portion of the spins may have been converted to orbital current at or near the interface. Further studies are required to quantify the charge-orbital-spin interconversion efficiency by combining different characterization techniques and taking into account the magnetic properties of specific materials.[55-57]

    In conclusion, we have demonstrated electrical manipulation of non-collinear AFM state in $Mn_3Sn$ by using 3d metal Cr as the spin current source. From combined current-

induced switching, angular-dependent MR and THz emission measurements in CoFeB/Cr, we can conclude that spin-charge interconversion in Cr is distinct from heavy metals and all experimental results converge to the OHE scenario. Our work provides an alternative way to electrically manipulate the spin states of non-colinear antiferromagnets and a method to disentangle SHE and OHE in light metals.

## AUTHOR INFORMATION

**Corresponding Author**

* Email: elewuyh@nus.edu.sg

**Author contributions**

Y.W. designed and supervised the project. H.X. and Y.W. designed the experiments. H.X. conducted the sample fabrication, electrical measurement, and magnetic characterization. N.Z. performed THz emission measurement. Y.M. helped with sample fabrication. X.C. performed XRD measurement. H.X., N.Z., and Y.W. analyzed the data. All authors discussed the results. H.X. and Y.W. wrote the manuscript and all the authors contributed to the final version of manuscript.

**Notes**

The authors declare no competing interests.

## ACKNOWLEDGEMENT

Y.W. acknowledge the funding supported by the Advanced Research and Technology Innovation Centre (ARTIC), the National University of Singapore under Grant No. A-0005947-09-00.